\documentclass{iucrjournals}

\usepackage{graphicx,pstricks}
\usepackage{graphics}
\usepackage{moreverb}
\usepackage{physics}
\usepackage{graphicx}
\usepackage{caption}
\usepackage{epsfig}
\usepackage{subcaption}
\usepackage{tikz}
\usepackage{amsmath}
\usepackage{dsfont}
\usepackage{amssymb}
\usepackage{array} 
\usepackage{makecell}
\usepackage{afterpage}
\usepackage[version=4]{mhchem}
\usepackage[T1]{fontenc} 

\newcommand{\mi}{\mathrm{i}} 

\graphicspath{{Figures/}}

\title{Resolving Tangling in Multi-Conformer Refinement via Iterative Projections}
 
\author[a]{Avinash Mandaiya\IUCrCemaillink{am2957@cornell.edu}\IUCrOrcidlink{0000-0002-4863-8244}}%
\author[a]{Veit Elser\IUCrEmaillink{ve10@cornell.edu}\IUCrOrcidlink{0000-0002-6961-4778}}

\affil[a]{Laboratory of Atomic and Solid State Physics, Cornell University, Ithaca, New York 14853-2501, USA}

\begin{document} 
\nolinenumbers
\maketitle 

\begin{synopsis}

Multi-conformer refinement frequently stalls due to clashes between geometric and density constraints, producing tangling among different conformations. Using an iterative method based on constraint projections, we are able to produce untangled refinements.
\end{synopsis}

\begin{abstract}
The advent of advanced crystallographic techniques has shifted structural biology from static, single-conformer models toward probing protein dynamics. Extracting cooperative motions from temporally and spatially averaged electron density maps requires both high-resolution data and refinement algorithms capable of handling conformational heterogeneity. However, current refinement protocols often fail due to the tangling phenomenon, in which conformational states become improperly intertwined during optimization. Here, we present an automated refinement methodology based on iterative projections within the divide-and-concur framework. This approach enables seamless integration of geometric constraints with experimental density constraints derived from observed scattering amplitudes. By allowing each atom to satisfy density constraints independently, we show that this framework effectively circumvents tangling artifacts and achieves robust refinement performance, even for models initialized with R-factors as high as 12\%. Just as iterative projections revolutionized phase retrieval in crystallography, we demonstrate that they can also address the optimization challenges in multi-conformational refinement. This work establishes a computational foundation for advancing crystallographic methodologies to resolve conformational heterogeneity and ultimately capture protein dynamics at atomic resolution.
\end{abstract}

\keywords{Multi-Conformer Refinement; X-ray Crystallography; Optimization; Iterative Projections}

\section{Introduction}

X-ray crystallography has remained the premier method for determining atomic-resolution structures of biological macromolecules, providing fundamental insights into protein structures and their function, drug-target interactions, and many other biological processes. Central to the success of structure determination using crystallography is a two-step process that addresses distinct computational challenges. Since crystallographic experiments measure only intensities (the magnitudes of the structure factors), the key challenge is determining the phases, a difficulty known as the phase problem. Resolving this problem is essential for reconstructing the real-space electron density at the experimental resolution, as the quality of the reconstruction depends critically on accurate phase information. For phase retrieval, iterative projection methods have proven highly effective \cite{Elser2003phase}, serving as a versatile framework that extends naturally to other non-convex constraint satisfaction problems \cite{elser2025solving} using the divide-and-concur framework \cite{gravel2008divide}. 

Once the phases are successfully retrieved and the real-space electron density map is obtained, an initial atomic model is constructed. This model then undergoes crystallographic refinement, during which the recovered phase information guides the model toward an optimal fit with the observed diffraction data while ensuring chemical and physical plausibility. The mathematical foundation of crystallographic refinement is based on minimizing the discrepancy between the observed structure factor amplitudes ($|F_{\text{obs}}|$) and those calculated from the atomic model ($|F_{\text{calc}}|$). This discrepancy is traditionally quantified using the $\text{R}$-factor,

\[
R = \frac{\displaystyle \sum_{\mathbf{k}} \left| |F_{\text{obs}}(\mathbf{k})| - |F_{\text{calc}}(\mathbf{k})| \right|}{\displaystyle \sum_\mathbf{k} |F_{\text{obs}}(\mathbf{k})|},
\]
where the sum is over all the measured reflections $\mathbf{k}$.

To assess model overfitting, a cross-validation metric known as the free $\text{R}$-factor ($R_{\text{free}}$) was introduced, which is calculated using a subset of reflections excluded from the refinement process \cite{Brunger1992free}.
 
Modern crystallographic refinement methods incorporate chemical constraints and prior knowledge to address experimental uncertainties and incomplete structural data. These advances have been implemented in widely used software packages, including \texttt{REFMAC5} \cite{Murshudov2011refmac5} and \texttt{PHENIX.refine} \cite{Afonine2012refine}, each offering unique strategies for optimizing model quality and agreement with the observed diffraction data through distinct mathematical frameworks. These refinement protocols typically employ restrained refinement strategies that combine the experimental data term with prior chemical knowledge through geometric restraints on bond lengths, angles, and dihedral angles. Specifically, \texttt{REFMAC5} employs maximum likelihood methods coupled with geometric restraints and incorporates TLS (Translation/Libration/Screw) refinement to model anisotropic displacements to capture rigid-body motions of protein domains. \texttt{PHENIX.refine} offers comprehensive refinement capabilities with automated optimization strategies and supports refinement across a broad range of resolution limits using X-ray, neutron, or both types of data simultaneously. In addition to atomic coordinates, these software packages also refine occupancy and anisotropic displacement parameters (ADPs).

The remarkable success of single-conformer structure determination has made crystallography a cornerstone of structural biology, with robust methodologies for both phase retrieval and refinement enabling reliable atomic-resolution models for well-diffracting crystals. Despite these methodological advances, significant challenges persist in crystallographic refinement, particularly when dealing with conformational heterogeneity. Traditional single-conformer refinement assumes a static, averaged structure, using isotropic or anisotropic B-factors to account for atomic motion and disorder. However, this approach becomes inadequate when proteins exist in multiple conformational states, as B-factors cannot distinguish between dynamic motion and static disorder \cite{Kuzmanic2014x}. 

The underlying data from X-ray diffraction represents temporally and spatially averaged structural information from large ensembles of protein molecules within the crystal lattice. When multiple alternative conformations coexist within these crystals, their individual electron density contributions become superimposed, resulting in weakened and averaged density maps that mask distinct structural states. This blending of conformational signals significantly complicates both the detection and accurate modeling of dynamic or functionally relevant protein states using traditional crystallographic refinement procedures.

A fundamental computational barrier has emerged in current refinement methods, called the \textit{tangling phenomenon}~\cite{Holton2025untangle}. Refinement programs like \texttt{REFMAC5}, \texttt{PHENIX.refine} etc., frequently become trapped in configurations where atoms are incorrectly swapped between alternative conformations, arising primarily from conflicts between the electron density term in the refinement target and prior constraints implied by chemistry. This tangling represents a core challenge when simultaneously trying to satisfy both experimental density data and stereochemical requirements in a multi-conformer refinement.

The development of robust methods for refining multiple conformational states, particularly those that can overcome the tangling phenomenon, represents a critical frontier in structural biology. As the field transitions from static structure determination toward characterizing protein dynamics through high-resolution crystallography, resolving conformational heterogeneity becomes essential for understanding how proteins function. This capability has direct implications for drug design, where conformational flexibility often governs binding specificity, and enzyme engineering, where catalytic mechanisms depend on dynamic structural transitions.

In this work, we present a divide-and-concur constraint satisfaction framework \cite{gravel2008divide} for the multi-conformational refinement problem, which is solved using an iterative projection method known as Reflect-Reflect-Relax (\texttt{RRR}). Just as iterative projection methods have proven transformative for phase retrieval, we show that they can similarly address the optimization challenges in multi-conformational refinement. We demonstrate that this approach circumvents the pitfalls associated with the tangling phenomenon with relative ease and can refine the system even when starting from an R-factor of 12\%. As we continue to push the boundaries of crystallographic methodology, addressing the complexities of conformational heterogeneity and the underlying optimization challenges through novel computational approaches remains essential for realizing the full potential of X-ray crystallography.

\section{Divide-and-Concur Framework}

The \textit{divide-and-concur} (D\&C) framework \cite{gravel2008divide} provides a general approach for solving constraint satisfaction problems through iterative projections. In the \textit{divide} part, the original problem is decomposed into a collection of smaller, more tractable subproblems, each defined by simpler constraints that can be efficiently solved using projection operators. To enable this decomposition, the framework introduces multiple replicas of each variable representing the system's state, with each replica responsible for only its assigned subproblem. The divide projection can be expressed as
\begin{equation}
    P_D(\mathbf{x}) = P_1(\mathbf{x^{(1)}}) \times P_2(\mathbf{x^{(2)}}) \cdots \times P_N(\mathbf{x^{(N)}}),
\end{equation}

where $P_i$ for $i \in \{1,2,\ldots,N\}$ represents the projection operator for the $i$-th subproblem, and $\mathbf{x}^{(i)}$ denotes the replica of variables assigned to that subproblem. For instance, maintaining the local rigid geometry of macromolecules involves enforcing constraints on bond lengths and bond angles, as well as other geometric restrictions, which can be implemented using independent replicas of the atomic positions. In the \textit{concur} part of the framework, these replicas are reconciled in a distance-minimizing manner to ensure mutual consistency. Additionally, when variables are interdependent, consistency must be maintained across them. For instance, in structure determination problems that utilize electron density maps, such as the one addressed in this work, the density variables must remain aligned with their corresponding atomic positions. The projection operator responsible for enforcing the concur constraint is denoted $P_C$ and is described in greater detail in the next section.

A feasible solution is obtained when the iterative search converges to a point that lies in both the divide and concur constraint sets. This search is performed using the Reflect--Reflect--Relax (\texttt{RRR}) algorithm, named after its iteration (update) rule,
\begin{equation}
x^\prime = \left(1-\frac{\beta}{2}\right)x + \frac{\beta}{2}R_C(R_D(x)) ,
\end{equation}
where $x$ is a general point in the space of all variables, $\beta$ is a ``time-step'' parameter restricted to the range $(0, 2)$, and $ R $ denotes the reflection operator associated with projection $P$ :
\begin{equation}
R(x) = 2\,P(x) - x.
\end{equation}
The iterative map can equivalently be written in terms of projection operators as
\begin{equation}\label{eq:RRR_update}
x^\prime = x + {\beta}\Big(P_C(2P_D(x) - x) - P_D(x)\Big) .
\end{equation}
The algorithm converges to a fixed point $x^*$ when  
\[
P_C\!\left( 2P_D(x^*) - x^* \right) = P_D(x^*),
\]  
indicating that a feasible solution satisfying both constraints has been obtained:  
\[
x_{\mathrm{sol}} = P_D(x^*) \in D \cap C.
\]
In the Untangle Challenge\footnote{\href{https://bl831.als.lbl.gov/~jamesh/challenge/twoconf/}{Untangle Challenge website: https://bl831.als.lbl.gov/~jamesh/challenge/twoconf/}}, two distinct types of constraints must be satisfied. The first consists of rigid-body constraints, which stem from local atomic geometry and are dictated by the quantum-mechanical hybridization states of the atoms. The second is the density constraint, imposed by the experimentally observed scattering data. We also need to ensure consistency between the model and the measured diffraction intensities. The following section describes a divide-and-concur framework specifically adapted to address these coupled constraints in the multi-conformer refinement problem.

\section{Divide Constraint Set and Projections}

Two fundamental classes of constraints govern refinement problems in structural biology --- geometric and density constraints. Geometric constraints preserve chemically valid local configurations by enforcing stereochemical arrangements, ensuring that candidate models remain physically realistic. Density constraints, on the other hand, require the refined atomic model to agree with the experimentally observed electron density, placing atoms where the electron density is most consistent with the experimental data. Accurate structural modeling demands that these two classes of constraints be satisfied simultaneously, balancing local geometry with fidelity to experimental data.

For this multi-conformer refinement problem, we assume that the phase information of the structure factors is known, as is often the case when refining from an already built atomic model. Under this assumption, the density constraint can be formulated in reciprocal space as a linear summation condition. For each Bragg reflection, the sum of atomic contributions must match the structure factor data. Because this density constraint is linear, it can be applied efficiently. In contrast, geometric constraints are inherently nonlinear and involve multi-body interactions, making them considerably more challenging to enforce.

Molecular geometry is determined by a combination of quantum-mechanical effects, such as orbital hybridization (e.g., $sp^2$, $sp^3$ geometries) and intermolecular forces, including van der Waals interactions, resonance stabilization, and atomic packing. These factors give rise to quasi-rigid structural motifs, such as peptide planes and side-chain aromatic rings, that resist deformation. Traditional force-field models capture these features by applying energy penalties to deviations from ideal bond lengths, bond angles, torsional angles, and improper dihedrals (out-of-plane bending). In our approach, these energy-based terms are reformulated as geometric constraints, allowing for more controlled and interpretable enforcement of molecular rigidity during refinement.

Geometric constraints vary in complexity. Pairwise constraints, such as bond lengths and volume exclusion, can be enforced with simple projection operations -- for example, by moving two atoms symmetrically on their common axis to the correct distance. Multi-body geometric constraints are more complex because they restrain the arrangements of three or more atoms. For instance, modifying a bond angle or torsion requires the coordinated repositioning of three or more atoms to achieve the correct angular arrangement, which cannot be accomplished through independent or sequential pairwise updates. To overcome this limitation, we use a motif-based approach that treats groups of atoms as rigid bodies of known structure.

A key challenge in a strict constraint-based approach is accommodating the small local distortions sometimes required for overall structural stability. To address this, we allow limited, controlled deviations from ideal motif geometry through a method known as \textit{short projection}. In this approach, the ideal motif configuration defines the center of a high-dimensional sphere, and the actual atomic configuration is constrained to lie within that sphere. The sphere’s radius specifies the maximum allowed deviation, effectively replacing the exact constraint with a small tolerance region. Projections are then performed with respect to this modified constraint set, preserving local rigidity while allowing the limited conformational flexibility needed for realistic macromolecular models to fit the electron density map.

The allowable deviation from an ideal motif geometry varies with both the nature of the constraint and the atoms it involves. Different constraint types have inherently different tolerances: deviations in bond lengths, for example, are typically penalized more strongly in energy functions than deviations from peptide bond planarity. Although planarity is structurally important, minor deviations are not only acceptable but often required, especially in flexible regions such as loops. Even within a single type of constraint, thresholds can differ depending on atomic composition. A dihedral angle defined by two hydrogen atoms may permit relatively large deviations because of their small radii and minimal steric hindrance, whereas constraints involving heavier atoms incur greater steric costs and therefore require stricter limits. We determine these tolerance values through empirical analysis of high-resolution protein structures, identifying statistically derived deviation ranges that maintain structural integrity while accommodating natural conformational variability.

Beyond the rigid-body constraints described above, our framework includes additional geometric constraints to address steric repulsion and specific covalent linkages involving cysteine residues. Steric clashes are avoided through a \textbf{volume exclusion constraint}, which ensures that non-bonded atomic pairs remain farther apart than a minimum distance that depends on the sum of their van der Waals radii. Covalent disulfide linkages are modeled through a \textbf{disulfide bond constraint}, which is applied when crystallographic evidence supports a bond between cysteine residues. The constraint identifies the correct cysteine pairs that form disulfide bonds and then enforces the ideal bond length and geometry. All geometric constraint types implemented in our framework are summarized below:

\subsection*{Geometric Constraint Types} 

\begin{itemize}

    \item \textbf{Rigid-Body Motifs} -- When a constraint involves more than two atoms, the corresponding atomic position variables are projected onto a rigid motif in the constraint set that preserves their internal geometry. These motifs are free to rotate and translate, adopting the orientation that minimizes the distance to the atomic positions. They can be categorized into:
    
    \begin{itemize}

        \item \textbf{Bond Angles} -- For three atoms (\ce{A-B-C}), this constraint enforces the ideal bond angle $\theta_0$ at the central atom by projecting the triplet onto a rigid motif with the correct angle.

        \item \textbf{Dihedral Angles} -- For four sequentially bonded atoms (\ce{A-B-C-D}), the allowed dihedral angles between the planes (\ce{A-B-C}) and (\ce{B-C-D}) are captured using a predefined set of motifs. Because dihedral energy profiles often exhibit multiple minima (e.g., gauche, trans), this set contains several permissible geometries. The configuration is projected onto the motif from this set that minimizes the distance to it.
        
        \item \textbf{Peptide Planarity} -- This constraint type ensures coplanarity in the backbone unit \ce{C_\alpha-CO-NH-C_\alpha} formed by peptide bonds between consecutive residues. Resonance between carbonyl and amide groups restricts \ce{C-N} bond rotation, producing a planar geometry essential for $\alpha$-helices and $\beta$-sheets.
        
        \item \textbf{Chirality and Rigid Groups} -- These preserve the correct stereochemistry at tetrahedral centers (e.g., \ce{C_\alpha} in amino acids) by projecting bonded groups onto the proper enantiomeric configuration, and treat chemically stable groups (e.g., aromatic rings, the guanidino group in arginine) as rigid units.
    
    \end{itemize}
    
    \item \textbf{Bond Lengths} -- Enforce a fixed distance between two bonded atoms equal to the equilibrium value  
    \[
    r_0 = \|\mathbf{r}_{\ce{A}} - \mathbf{r}_{\ce{B}}\|.
    \]  
    This is achieved by scaling the separation vector between the atoms.
    
    \item \textbf{Volume Exclusion} -- These prevent steric clashes by enforcing a minimum separation between non-bonded atoms, as defined by the \textit{MolProbity} \cite{chen2010molprobity,word1999asparagine} criterion.

    \item \textbf{Disulfide Bonds} — When crystallographic evidence indicates the presence of disulfide bonds, the cysteine residues are paired in a distance-minimizing way during the iterative search.    
\end{itemize}

\subsection{Geometric Constraints and Projections}

Enforcing geometric constraints requires projecting atomic configurations onto sets defined by each constraint type. In this section, we mathematically formulate the constraint sets and then describe projection methods tailored to each constraint class.

\subsubsection{Bond Length Constraint}
The bond length constraint acts on pairwise interatomic distances. For covalently bonded atoms $A$ and $B$ with an equilibrium bond length $b_0$ and a tolerance $\delta b$, the constraint set is defined as
\begin{equation}
\mathcal{C}_{\text{bond}} \colon \left| \, \left\|\mathbf{r}_A - \mathbf{r}_B\right\| - b_0 \, \right| \leq \delta b
\end{equation}
The projection onto the bond length constraint set is straightforward to compute in closed form, even when a permissible deviation range is allowed. Given initial positions $(\mathbf{r}_{A}^{(0)}, \mathbf{r}_{B}^{(0)})$, the bond length projection operator maps them to the nearest configuration whose separation lies within the tolerance window $[b_0^-,\, b_0^+]$, where
\[
b_0^\pm = b_0 \pm \delta b.
\]
The projection is given by:
\begin{equation}
P_{\mathcal{C}_{\text{bond}}}(\mathbf{r}_{A}^{(0)}, \mathbf{r}_{B}^{(0)}) = 
\begin{cases}
\left( \mathbf{r}_{A}^{(0)} + \lambda^- \mathbf{u},\ \mathbf{r}_{B}^{(0)} - \lambda^-\mathbf{u} \right) & \text{if } \big\|\mathbf{d}\big\| < b_0^- \\
\left( \mathbf{r}_{A}^{(0)} + \lambda^+\mathbf{u},\ \mathbf{r}_{B}^{(0)} - \lambda^+\mathbf{u} \right) & \text{if } \big\|\mathbf{d}\big\| > b_0^+ \\
\left( \mathbf{r}_{A}^{(0)},\ \mathbf{r}_{B}^{(0)} \right) & \text{otherwise}
\end{cases}
\end{equation}

where
\begin{itemize}
\item $\mathbf{d} = \mathbf{r}_{A}^{(0)} - \mathbf{r}_{B}^{(0)}$ is the initial interatomic separation vector,
\item $\mathbf{u} = \mathbf{d}/\|\mathbf{d}\|$ denotes the unit vector along the bond axis, and 
\item $\lambda^\pm = \frac{1}{2}(\|\mathbf{d}\| - b_0^\pm)$ quantifies the required displacement magnitude.
\end{itemize}

\subsubsection{Rigid Body Constraint}

A rigid body constraint applies when a group of $N$ atoms must preserve their internal geometry. Let $\mathbf{M} \in \mathbb{R}^{N \times 3}$ denote the reference geometry motif containing the equilibrium positions of the $N$ atoms involved. Valid configurations under this constraint are given by the constraint set,  
\begin{equation}
  \mathcal{C}_{\text{rigid}} \colon \mathbf{R} = \mathbf{MU} + \mathbf{T},  
  \quad \text{where} \quad
  \begin{aligned}
    &\mathbf{U} \in \mathbb{SO}(3) \quad &&\text{(rotation matrix)}, \\  
    &\mathbf{T} = \mathbf{1}_N \otimes \mathbf{t}^\intercal \in \mathbb{R}^{N \times 3} \quad &&\text{(row-wise translation)}.  
  \end{aligned}
\end{equation}  
In more detail,
\begin{itemize}
\item $\mathbf{M} = [\mathbf{m}_1, \mathbf{m}_2, \dots, \mathbf{m}_N]^\intercal$ contains the reference positions of the motif in its equilibrium geometry. The number of distinct motifs required depends on the constraint type. For example, a bond angle constraint needs only one motif, whereas a planar peptide constraint may require multiple motifs (e.g., both \textit{cis} and \textit{trans} configurations for a proline residue).
\item $\mathbf{U}$ applies a rigid-body rotation, preserving all internal bond lengths and angles.
\item $\mathbf{T}$ applies the same translation vector $\mathbf{t} \in \mathbb{R}^3$ to every atom in the motif.
\end{itemize}

Projection onto $\mathcal{C}_{\text{rigid}}$ is carried out using the Kabsch–Umeyama algorithm \cite{kabsch1976solution, umeyama2002least}, which determines the optimal rotation and translation that aligns the configuration with the nearest valid rigid-body geometry. 

\subsubsection{Volume Exclusion Constraint}

To maintain physical plausibility, atoms in a molecular structure must not overlap. The volume exclusion constraint enforces this by preventing configurations in which the distance between atomic centers is less than the sum of their van der Waals radii, adjusted by a tolerance parameter. This eliminates steric clashes and promotes realistic packing, which is essential for chemically and physically viable conformations.

For two non-bonded atoms $A$ and $B$ with van der Waals radii $r_A$ and $r_B$, the minimum allowed center-to-center distance is
\[
r_0 = r_A + r_B - t_0 ,
\]
where $t_0$ is the tolerance. By default, $t_0 = 0.4 \, \text{\AA}$ following the MolProbity criterion, although it may vary under specific conditions. For example, hydrogen-bonding pairs are assigned a larger tolerance to accommodate closer contacts.

The corresponding constraint set is defined as
\begin{equation}
\mathcal{C}_{\text{ex}} \colon  \|\mathbf{r}_A - \mathbf{r}_B\| > r_0 
\end{equation}

The projection to the volume exclusion constraint is achieved by moving the two atoms symmetrically away from each other along the direction $\mathbf{d} = \mathbf{r}_{A}^{(0)} - \mathbf{r}_{B}^{(0)}$ until they are separated by distance $r_0$. 
\begin{equation}
P_{\mathcal{C}_{\text{ex}}}(\mathbf{r}_{A}^{(0)}, \mathbf{r}_{B}^{(0)}) = 
\begin{cases}
\left( \mathbf{r}_{A}^{(0)} +  \frac{r_0 - \| \mathbf{d}\|}{2\| \mathbf{d}\|}\mathbf{d}\ , \ \mathbf{r}_{B}^{(0)} - \frac{r_0 - \| \mathbf{d}\|}{2\| \mathbf{d}\|}\mathbf{d} \right) & \text{if } \big\|\mathbf{d}\big\| < r_0 \\
\left( \mathbf{r}_{A}^{(0)},\ \mathbf{r}_{B}^{(0)} \right) & \text{otherwise.}
\end{cases}
\end{equation}
\subsubsection{Disulfide Bond Constraint}

To identify which cysteine residues form disulfide bonds, we formulate a matching constraint that enforces one-to-one bonding between distinct cysteines. Given $N_S$ cysteine residues in the protein, there are $N_S(N_S-1)/2$ possible unordered pairs that could potentially form a disulfide bond. However, each cysteine can participate in only one such bond. To convert this problem into a matching problem, we transform the unordered pairs to ordered pairs such that each cysteine residue is paired exactly once but appears twice in the perfect matching set as $(i,j)$ and $(j,i)$ due to the symmetry. This is achieved via a minimum-cost matching algorithm that selects the most geometrically compatible pairs.

The cost matrix $\Delta$ encodes geometric deviation from the ideal bond length. Specifically, the entry $\Delta_{ij}$ quantifies the squared distance between the unprojected positions of sulfur atoms $i$ and $j$ to the disulfide bond length constraint. For a specified tolerance window $[b_0^-, b_0^+]$, the cost is given by:
\begin{align}
\Delta_{ij}^2 = 
\begin{cases} 
\frac{1}{2}\left( \|\mathbf{r}_i^{(0)} - \mathbf{r}_j^{(0)}\| - b_0^- \right)^2, & \|\mathbf{r}_i^{(0)} - \mathbf{r}_j^{(0)}\| < b_0^- \\
0, & b_0^- \leq \|\mathbf{r}_i^{(0)} - \mathbf{r}_j^{(0)}\| \leq b_0^+ \\
\frac{1}{2}\left( \|\mathbf{r}_i^{(0)} - \mathbf{r}_j^{(0)}\| - b_0^+ \right)^2, & \|\mathbf{r}_i^{(0)} - \mathbf{r}_j^{(0)}\| > b_0^+
\end{cases}
\end{align}

where $\mathbf{r}_i^{(0)}$ and $\mathbf{r}_j^{(0)}$ are the initial (unprojected) positions of sulfur atoms $i$ and $j$. The optimal set of pairings is obtained by minimizing the total cost over all perfect matchings. A perfect matching $Q$ can be represented by a square $0/1$ permutation matrix $P$. The optimal distance-minimizing set of disulfide bonds $\mathcal{S}$ is determined using the Hungarian algorithm \cite{kuhn1955hungarian},
\begin{align}
\mathcal{S} = \left\{ (i, j) \ \big| \ P^*_{ij} = 1 \right\}, \quad \text{where} \quad P^* = \underset{P \in \mathcal{P}}{\arg\min} \ \mathrm{trace}(P\Delta),
\end{align}

where $\mathcal{P}$ denotes the set of all permutation matrices of size $N_S$. The resulting set $\mathcal{S}$ specifies the cysteine residue pairs that are constrained to satisfy disulfide bond geometry by projection onto the bond length constraint.

\subsection{Structure Factor Constraint and Projection}

The structure factor constraint enforces that the calculated, phase-resolved structure factor equals the sum of scattering contributions from all atoms in the model. For multi-conformer refinement, the contribution of each atom is scaled by the occupancy probability associated with its conformation. Standard refinement protocols (e.g., \textsc{Phenix} or \textsc{Refmac}) treat these occupancies as adjustable parameters, updating them iteratively to maximize agreement with the observed structure factors $F_{\mathrm{obs}}$. 

In the divide-and-concur framework, this can also be formulated by introducing occupancy probability variables constrained to sum to unity for each atom. However, because the focus of this study is on resolving tangling artifacts that frequently arise in traditional refinement protocols, we set the occupancies fixed throughout the refinement process rather than updating them iteratively. The fixed values are chosen from the ground-truth model, allowing us to isolate the effect of our projection-based strategy on positional refinement. 

For the protein model used in this study (PDB ID: \textbf{1AHO}), each atom is modeled with two alternative conformations, each assigned an occupancy of 0.5. While determining the correct occupancies of solvent molecules is more challenging due to weaker density signals, we apply the same fixed occupancy assumption to solvent atoms for consistency.

Let $\mathbf{k}$ denote a scattering vector for which the structure factor is experimentally determined, and let $\mathcal{I}$ denote the index set of all atom-conformation pairs. Then the structure factor constraint is expressed as:
\begin{equation}\label{eq:density_sum_constraint}
\mathcal{R}: \hspace{1cm} \sum_{c, a \in \mathcal{I}} p_{c a} \, \rho_{c a}[\mathbf{k}] = F_o[\mathbf{k}].
\end{equation}

Here, $p_{c a}$ is the occupancy of atom $a$ in conformation $c$, $\rho_{c a}[\mathbf{k}]$ is the contribution of that atom to the structure factor of conformer $c$ at scattering vector $\mathbf{k}$, and $F_o[\mathbf{k}]$ is the experimentally observed structure factor. This linear constraint must be satisfied for all the measured scattering vectors.

Let $\boldsymbol{\rho}^{(0)}(\mathbf{k}) \in \mathbb{C}^{|\mathcal{I}|}$ denote the vector of unprojected structure factors from all atom–conformation pairs at scattering vector $\mathbf{k}$. The objective is to determine the projected vector $\boldsymbol{\rho}(\mathbf{k})$ that satisfies the linear constraint in Equation~\ref{eq:density_sum_constraint} while minimizing the squared $\ell_2$-norm distance to $\boldsymbol{\rho}^{(0)}(\mathbf{k})$.
This leads to the constrained optimization problem
\[
\min_{\boldsymbol{\rho}(\mathbf{k})} 
\sum_{c,a \in \mathcal{I}} 
\left\lVert \rho_{c,a}(\mathbf{k}) - \rho^{(0)}_{c,a}(\mathbf{k}) \right\rVert^2
\quad \text{subject to} \quad
\sum_{c,a \in \mathcal{I}} p_{c,a} \, \rho_{c,a}(\mathbf{k}) = F_o(\mathbf{k}).
\]
This is solved using the method of Lagrange multipliers, resulting in the projected densities 
\begin{equation}\label{eq:proj_density}
\rho_{c,a}(\mathbf{k}) = \rho_{c,a}^{(0)}(\mathbf{k}) - \frac{p_{c,a}}{\displaystyle\sum_{c, a \in \mathcal{I}} p_{c,a}^2}\left(\sum_{c, a \in \mathcal{I}} p_{c,a} \, \rho_{c,a}^{(0)}(\mathbf{k}) - F_o(\mathbf{k})\right).
\end{equation}

\section{Concur Constraint and Projection}
In the divide-and-concur framework, the concur step enforces consistency across variables that are subject to multiple constraints. For the refinement problem, atomic positions that participate in different geometric constraints are represented by separate replicas. These replicas permit geometric constraints to be satisfied independently, but they must ultimately be aligned to represent a single physically consistent atomic position. The concur constraint set enforces this alignment, requiring all replicas of the same atom to coincide in space.

In addition to geometric variables, the concur constraint also applies to electron density variables, which depend explicitly on the atomic coordinates. The contribution of an atom to the electron density in reciprocal space can be expressed as
\[
\rho(\mathbf{k}; \mathbf{r}) = \rho(\mathbf{k}; 0)\exp(\mi\mathbf{k} \cdot \mathbf{r}),
\]
where $\rho(\mathbf{k}; 0)$ is the atomic form factor, representing the reciprocal-space density of the atom when located at the origin. This form factor is well-approximated by a weighted sum of Gaussians,
\begin{align}\label{eq:gaussian_density}
\rho(\mathbf{k}; 0) = \sum_{i=1}^{4} a_i \exp\left( -b_i \left( \frac{|\mathbf{k}|}{4\pi} \right)^2 \right) + c,
\end{align}
where the constants $a_i$, $b_i$, and $c$ depend on the atomic type.

\begin{figure}[t]
    \centering
    \includegraphics[width=0.48\linewidth]{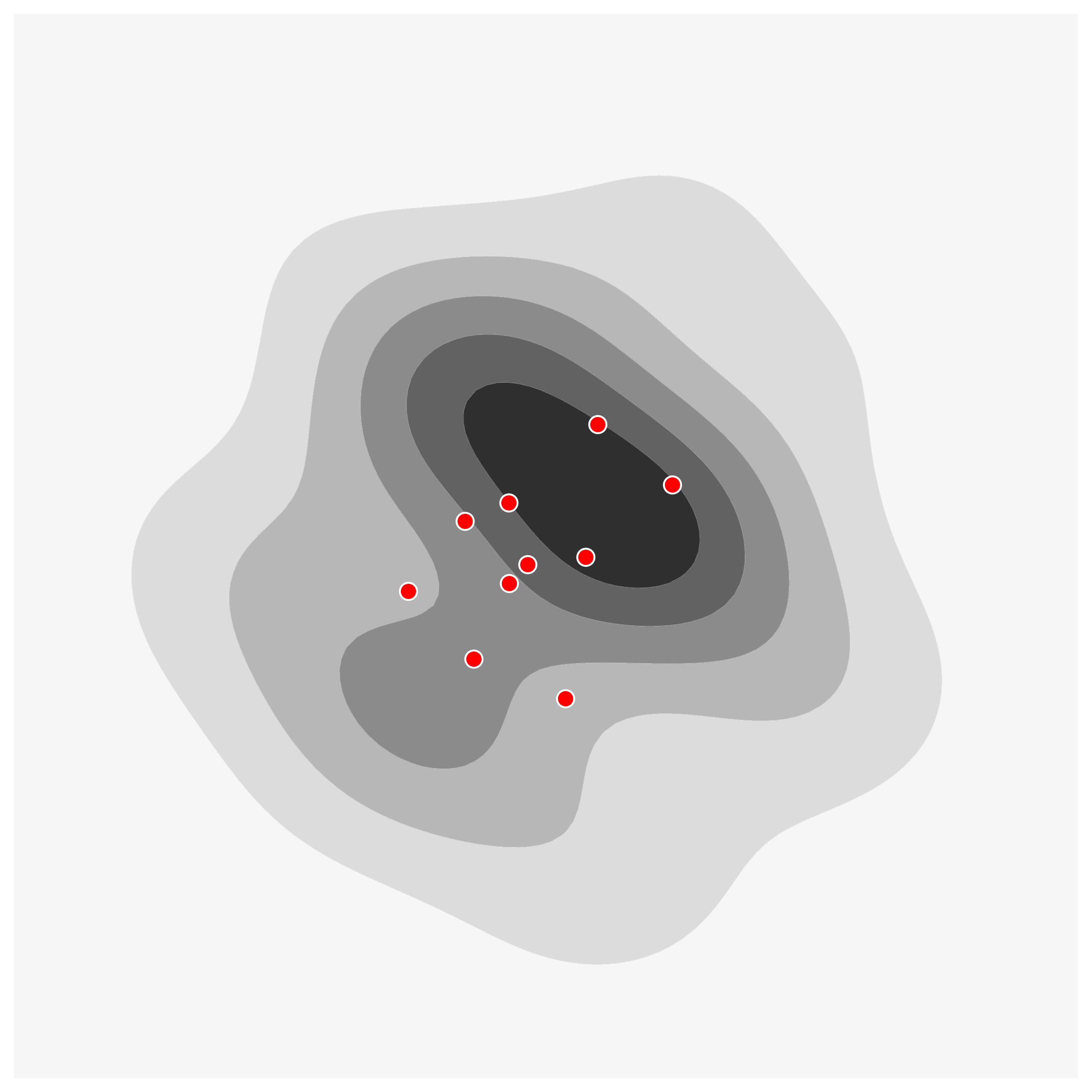}
    \hfill
    \includegraphics[width=0.48\linewidth]{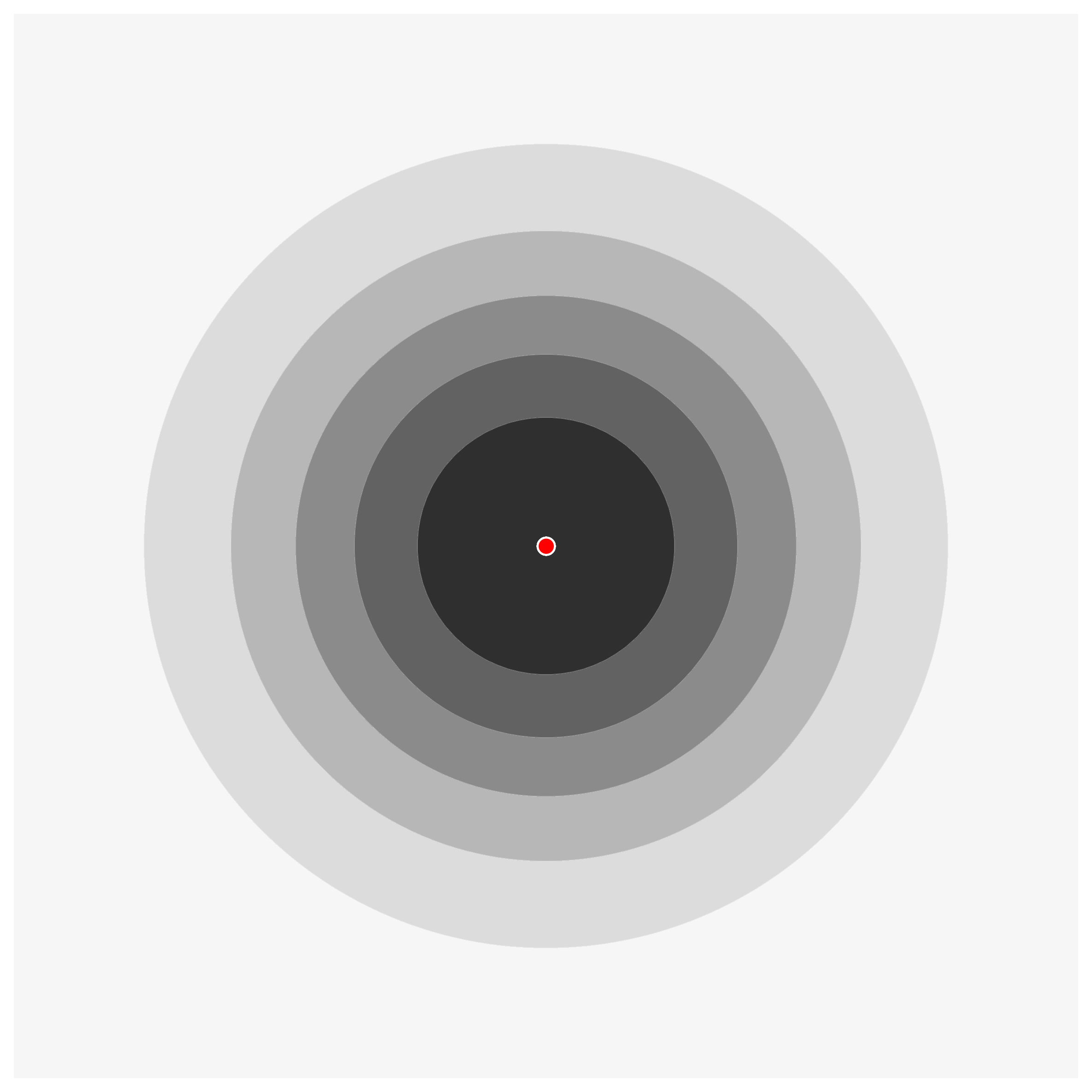}
    \caption{Cartoon illustration of the concur projection. \textbf{Left:} Multiple position replicas representing the same atom are initially distinct, accompanied by distorted electron density variables. \textbf{Right:} After the concur projection, the position replicas are reconciled with one another and made consistent with the density variables.}
    \label{fig:concur_density}
\end{figure}

Let $\mathbf{K}$ be the set of scattering vectors, and let $\mathbf{r}_d$ denote the position replica associated with constraint $d \in \mathcal{D}$. Here, $\mathcal{D}$ denotes the full set of constraints that are independently enforced in the divide set.

The concur constraint set is defined as:

\begin{equation}\label{eq:concur_constraint}
C: \quad \left\{
\begin{aligned}
&\rho(\mathbf{k};\mathbf{r}) = \rho(\mathbf{k} ; 0) \exp(\mi\mathbf{k} \cdot \mathbf{r}), \quad \forall \;\mathbf{k} \in \mathbf{K}, \\
&\mathbf{r}_d = \mathbf{r}, \quad \forall \; d \in \mathcal{D}.
\end{aligned}
\right.
\end{equation}

Here, the first condition ensures consistency between the electron density variables and the atomic position $\mathbf{r}$, while the second ensures that all atomic position replicas are consistent with each other.

The projection onto the concur constraint is determined by minimizing the squared distance
\begin{equation}\label{eq:distance_mapping_replicas}
\Delta^2(\mathbf{r};  \mathbf{r}^{(0)},\rho^{(0)}(\mathbf{k})) =
\sum_{d \in \mathcal{D}} \eta_d^2 \, \norm{\mathbf{r} - \mathbf{r}_d^{(0)}}^2
+ g^2 \sum_{\mathbf{k} \in \mathbf{K}} \eta_{k}^2 \,
\norm{\rho(\mathbf{k} ; 0) \exp(\mi\mathbf{k} \cdot \mathbf{r}) - \rho^{(0)}(\mathbf{k})}^2,
\end{equation}

where $\mathbf{r}^{(0)}$, given by $\{\mathbf{r}^{(0)}_d\}_{d \in \mathcal{D}}$, and $\rho^{(0)}(\mathbf{k})$ are the current iterates of the positions and densities, respectively, and $\eta_d$, $g \,\eta_{k}$, are constraint-specific metric parameters. The first term penalizes deviations from the position replicas, while the second penalizes mismatches between the density implied by $\mathbf{r}$ and the current density variables.

The standard isotropic Euclidean metric is replaced with a constraint-weighted metric parameterized by $\eta_d$ for geometric constraints and $g\,\eta_{\mathbf{k}}$ for density constraints. These parameters preserve the local convergence properties of the \texttt{RRR} algorithm while providing enhanced flexibility for handling near-solutions, where progress may stall. They are particularly valuable in tangled refinement scenarios, where certain types of constraints may dominate during the iterative search and hinder convergence. For instance, the volume-exclusion constraint introduces replicas for every atomic pair, even though most atoms are far apart, making many of these constraints trivial. Because treating such constraints on equal footing is not optimal, we assign weights explicitly: for all volume-exclusion constraints we set $\eta_d = 1$; for other geometric constraints we use $\eta_d = 10$, and for density constraints we set $\eta_{\mathbf{k}} = 0$ when $\lVert \mathbf{k} \rVert < \bar{k}$. The exclusion of low-$\mathbf{k}$ terms reflects the fact that small-angle scattering is particularly susceptible to systematic errors and background contributions, which introduce significant noise and make these measurements unreliable for guiding refinement. Here we use $\bar{k} = 0.4 \times 2\pi \,\text{\AA}^{-1}$ as the cutoff. For all remaining Bragg peaks, we assign $g\,\eta_{\mathbf{k}} = 0.1$.

The relative scaling parameter $g$ is used to balance the contributions of (and restore common units to) the two variable types -- atomic positions and electron density -- in the distance function. This ensures that both contribute on a comparable scale, preventing either from dominating the refinement process.

The non-linearity in the distance function \eqref {eq:distance_mapping_replicas} presents a significant challenge in the refinement problem, since an exact projection that minimizes the distance cannot be expressed in closed form. To address this, we employ an iterative tangent-space approximation scheme to obtain a locally distance-minimizing projection that enforces the constraint. We expect this local minimum to be the same as the global minimum. Further details on this method are provided in Appendix~\ref{app:mapping}.

\begin{figure}[t]
    \centering
    \begin{tikzpicture}
        \node[inner sep=0pt] (leftimg) at (0,0) {\includegraphics[width=0.48\linewidth]{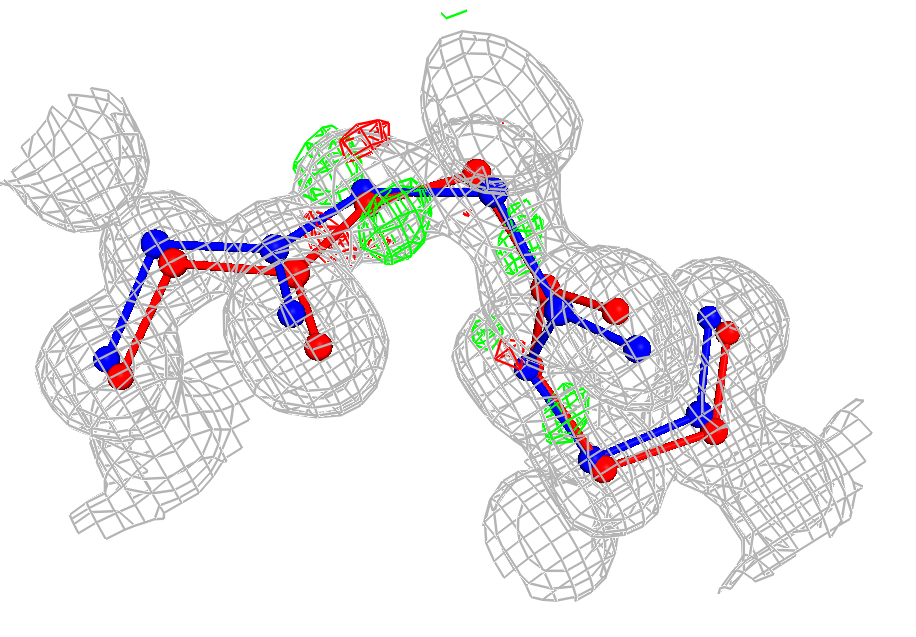}};
        \node[inner sep=0pt, anchor=west] (rightimg) at ([xshift=4mm]leftimg.east) {\includegraphics[width=0.48\linewidth]{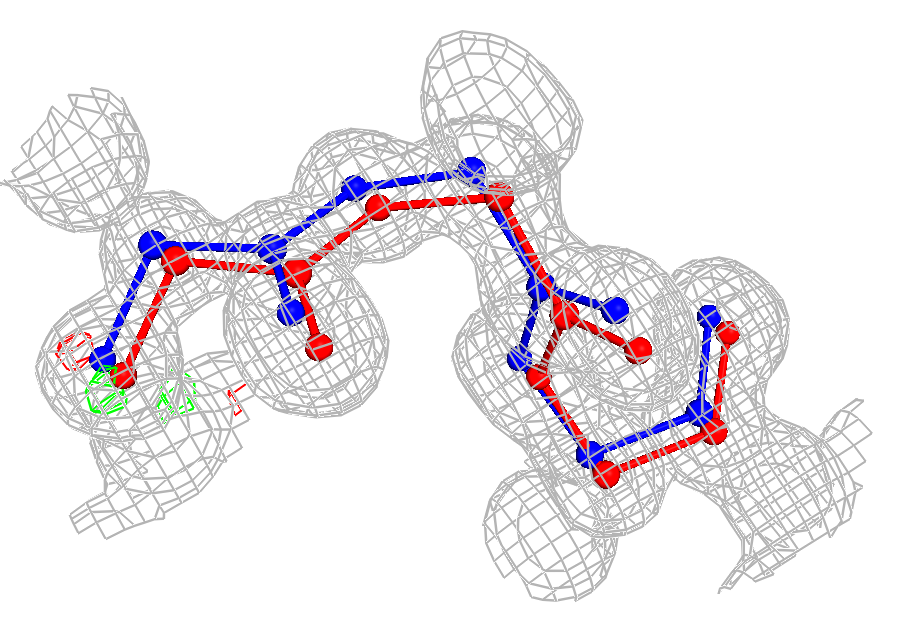}};

        \draw[->, black, very thick] (0.3, 2.5) -- (0.2, 1.4) node[midway, above left, xshift=-2pt, yshift=2pt, text=black] {\footnotesize swapped $C_\alpha$ atoms};
    \end{tikzpicture}
    \caption{Comparison of electron density maps before and after refinement for model \texttt{otw39.pdb}. The red and green meshes represent negative and positive deviations, respectively, from the optimal fit to the diffraction data, highlighting regions of poor agreement. \textbf{Left}: Electron density of the initial trapped configuration, where the \ce{C_\alpha} atom of residue 39 is incorrectly swapped between the two conformers. \textbf{Right}: Electron density following refinement using the iterative projection protocol with the \texttt{RRR} algorithm, demonstrating successful resolution of the conformational tangling.}
    \label{fig:L1_comparison}
\end{figure}

\section{Results}

\begin{figure}[htp]
    \centering

    \begin{subfigure}[t]{\linewidth}
        \centering
        \includegraphics[width=\linewidth]{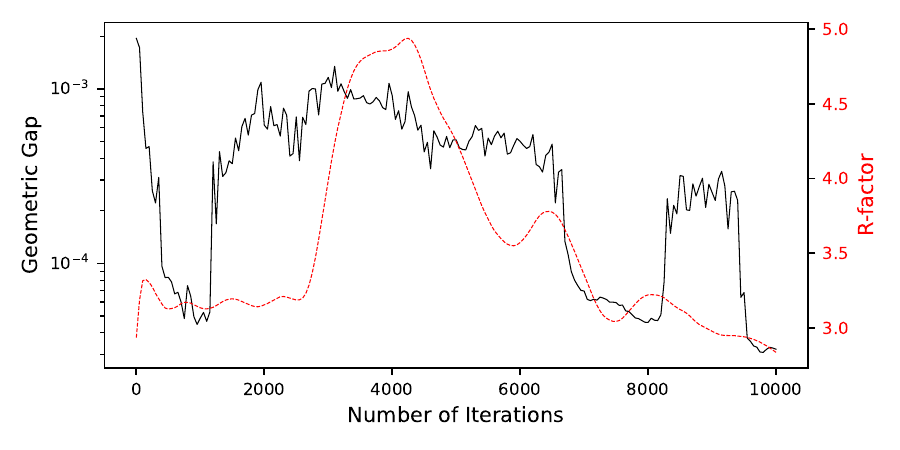}
        \label{fig:gapL1a}
    \end{subfigure}
      
    \begin{subfigure}[t]{\linewidth}
        \centering
        \includegraphics[width=\linewidth]{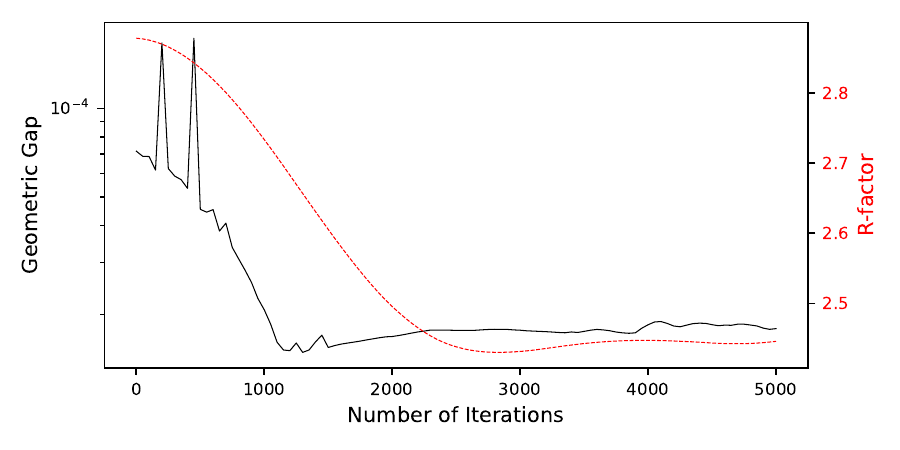}
        \label{fig:gapL1b}
    \end{subfigure}
    
    \caption{\textbf{Top}: RRR run for \texttt{otw39.pdb} showing the evolution of the R-factor (red line) and the constraint violation gap between the divide and concur constraint sets throughout the iterative refinement process. The constraint gap specifically measures violations in geometric constraints. \textbf{Bottom}: The final structure obtained from the previous run is further refined against the density data to achieve a better fit.}
    \label{fig:gap_L1}
\end{figure}

Figure~\ref{fig:L1_comparison} shows the initial atomic model (\texttt{otw39.pdb}), which becomes trapped in a configuration where the \ce{C_\alpha} atom of residue 39 is incorrectly swapped between two conformers. This tangled state produces a geometric outlier in the \ce{N-C_\alpha} bond of the alanine residue. The corresponding refinement trajectory, obtained using the \texttt{RRR} algorithm, is presented in Figure~\ref{fig:gap_L1}.

\begin{figure}[t]
    \centering
    \includegraphics[width=0.48\linewidth]{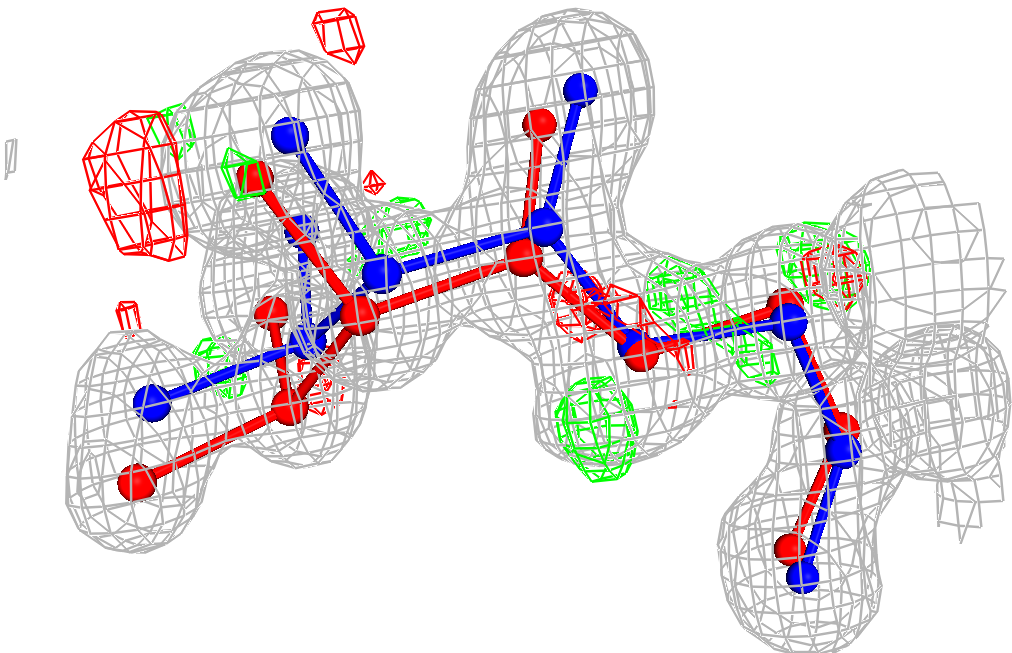}
    \hfill
    \includegraphics[width=0.48\linewidth]{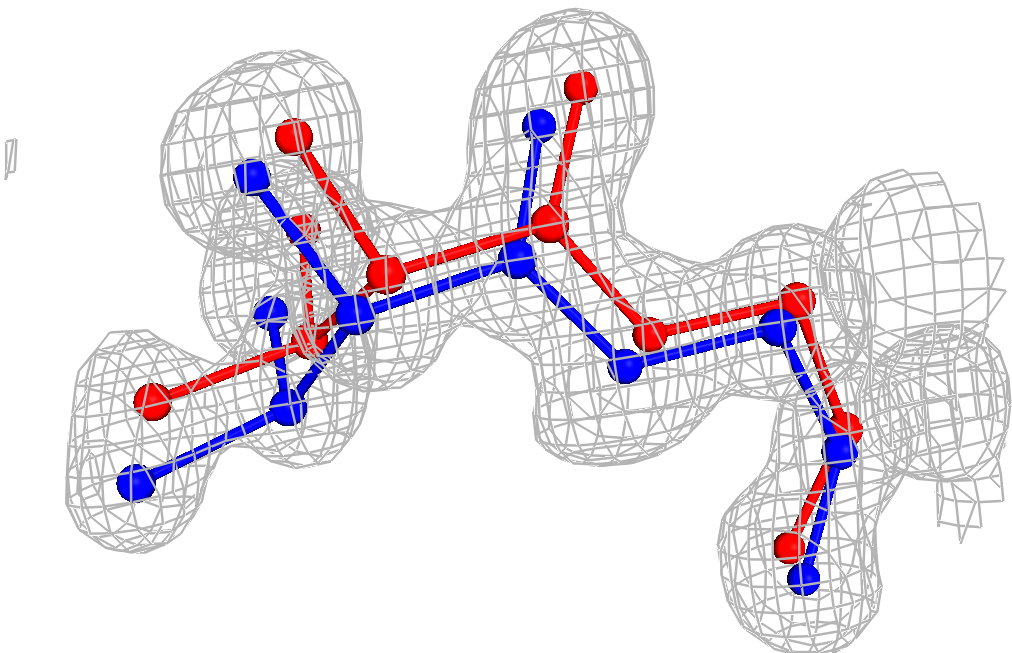}

    \caption{Electron density maps illustrating the refinement process for model \texttt{otw1.pdb}. \textbf{Left}: Initial configuration with the density map showing the entire side chain of the Valine residue incorrectly swapped due to conformational tangling. \textbf{Right}: Refined density obtained through the iterative projection protocol with the \texttt{RRR} algorithm.}
    \label{fig:L2_comparison}
\end{figure}

Despite the geometric error, the model exhibits a low initial $\text{R}$-factor of 2.9\%, indicating seemingly good agreement with the electron density. This misleading agreement contributes to the persistence of the tangled \ce{C_\alpha} positions in conventional refinement procedures. Using the \texttt{RRR} algorithm, approximately 10,000 iterations are required to escape the trapped configuration and achieve a state in which both the density and geometric constraints are satisfied. 

Although geometric constraints are crucial for guiding the model toward physically plausible structures, they can also impose overly restrictive conditions that hinder convergence. After the first \texttt{RRR} refinement, the initial tangled configuration is resolved, as seen in the top panel of Figure~\ref{fig:gap_L1}, with both geometric and density constraints mutually satisfied at an $\text{R}$-factor of roughly 2.9\%. At this stage, re-initializing the density variables -- based on the atomic positions obtained at the end of the first run -- proves substantially more effective for reaching a sub-2.5\% $\text{R}$-factor than simply increasing the iteration count in the initial run. This dual-stage refinement protocol, which incorporates re-initialization near the solution, enhances convergence and is particularly advantageous in noisy systems where the constraint set does not admit an exact solution.

We further assess the method on a more challenging case involving complete side-chain swapping between conformations in model \texttt{otw1.pdb}, as shown in Figure~\ref{fig:L2_comparison}. In the initial configuration, the entire side chain of the valine (\texttt{Val1}) residue is incorrectly exchanged between the two conformational states, trapping the model in a local minimum. Using the dual-stage refinement protocol, the \texttt{RRR} algorithm successfully resolves this tangled configuration.

Table~\ref{tab:rfactor_table} reports the $\text{R}$-factor for the initially tangled configurations of both models, as well as the results after the first and second stages of the dual-stage refinement protocol. For all the refinement runs, $\beta$ in the \texttt{RRR} update rule \eqref {eq:RRR_update} is set to be $0.5$. Additionally, the scaling parameter $g$ introduced in the combined position-density distance function \eqref {eq:distance_mapping_replicas} is set to be $0.1$. 

\begin{table}[ht]
\caption{R-factor values at different stages of refinement using the \texttt{RRR} algorithm.} 
\smallskip
\begin{center}
\begin{tabular}{llcr}
\midrule
\textbf{Model} & \textbf{Initial $\text{R}$-factor} & \textbf{After 1st Refinement} & \textbf{After 2nd Refinement} \\
\midrule
\texttt{otw39.pdb} & 2.9345 & 2.8775 & 2.4464 \\
\texttt{otw1.pdb}  & 2.8991  & 2.6989   & 2.3819 \\
\end{tabular}
\end{center}
\label{tab:rfactor_table}
\end{table}

\begin{figure}[htp]
    \centering
    \includegraphics[width=\linewidth]{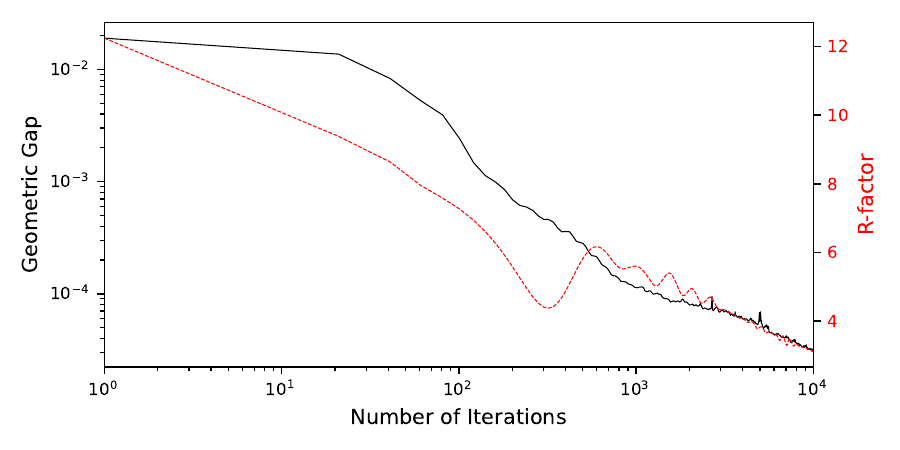}
    \caption{Convergence behavior of the \texttt{RRR} algorithm near the ground truth model for the protein (PDB ID: \textbf{1AHO}). The geometric gap evolution is shown for a refinement starting from all ground-truth atomic positions perturbed by displacements in the range $[-0.1,0.1]$\,\AA. The plot illustrates the systematic reduction of geometric constraint violations over iterations, with the correlated decrease in the $\text{R}$-factor.}
    \label{fig:p1_gap}
\end{figure}

Figure~\ref{fig:p1_gap} demonstrates convergence testing in the vicinity of the ground truth model for the protein (PDB ID: \textbf{1AHO}). The iterate is initialized by perturbing each atom's position from the ground truth model by a displacement randomly sampled from the uniform distribution in $[-0.1\,\mbox{\AA},0.1\,\mbox{\AA}]^3$ . The \texttt{RRR} algorithm demonstrates robust performance, successfully refining the structure even when starting from a state with an $\text{R}$-factor of 12\%. As the iterate approaches the true solution, we observe a strong correlation between the multiplicative changes in the geometric gap and the linear changes in the $\text{R}$-factor.

\section{Discussion}

In this work, we have demonstrated that iterative projection methods, exemplified by the \texttt{RRR} algorithm, can be applied with high effectiveness to the multi-conformer refinement problem. By formulating the refinement as a constraint satisfaction problem using a divide-and-concur framework, we introduce a structured approach to navigating the strongly non-convex optimization landscape. This framework allows geometric and density constraints to be enforced independently within the divide set, ensuring that each constraint is satisfied without compromising the other.

Conventional refinement methods typically maximize a log-likelihood target by adjusting all parameters simultaneously. Although often effective, this strategy is susceptible to trapping in local minima, which can result in conformational tangling for multi-state models. In contrast, our projection-based approach enforces constraint sets independently through replicas, allowing the algorithm to resolve tangled states more effectively.

The results highlight the method’s ability to resolve such tangled states in an automated manner. Most notably, \texttt{RRR} achieves sub-2.5\% $\text{R}$-factors while successfully disentangling side-chain and backbone conformations that have resisted conventional refinement procedures. Furthermore, the algorithm exhibits robust convergence even from highly perturbed starting models, with initial $\text{R}$-factors as high as 12\%, consistently recovering accurate, low-$\text{R}$ solutions. It is quite likely that when the entire refinement is done using the constraint-based method, the tangling phenomenon does not occur.

In conclusion, the implementation of iterative projections methodology to multi-conformer refinement represents a significant step forward in addressing one of the most challenging aspects of modern crystallographic analysis. As structural biology continues to grapple with increasingly complex systems exhibiting conformational heterogeneity, computational methods that can reliably navigate the associated optimization challenges become essential tools for realizing the full potential of experimental structural data. The framework presented here provides a robust foundation for future developments in this critical area of structural biology.
     
\appendix %

\section{Concur Constraint with Atomic Position Replicas and Density Variables}
\label{app:mapping}

The exponential dependence of the density term on $\mathbf{r}$ in \eqref{eq:distance_mapping_replicas} precludes closed-form solutions. We therefore employ an iterative scheme based on the tangent-space approximation. At each step, the non-linear term is linearized around the current estimate of $\mathbf{r}$, reducing the problem to a tractable form that can be solved efficiently. This strategy, previously shown effective in non-convex constraint satisfaction problems such as non-negative matrix factorization~\cite{elser2016tangent}, enables rapid convergence to the true projection. 

\subsection{Concur projection via iterative tangent space approximations}

The iterative tangent-space approximation provides a practical alternative to find the projection to the concur constraint efficiently. The key idea is to replace the non-linear factor $\exp(i \mathbf{k} \cdot \mathbf{r})$ with its first-order expansion about a reference point $\mathbf{r}^*$:  
\begin{align*}
    \exp(i \mathbf{k} \cdot \mathbf{r}) 
    &\approx \exp(i \mathbf{k} \cdot \mathbf{r}^*) 
    \Big(1 + i \mathbf{k} \cdot (\mathbf{r} - \mathbf{r}^*)\Big).
\end{align*}

This substitution reduces the exponential to a linear function of the deviation from the reference point $\boldsymbol{\delta r} = \mathbf{r} - \mathbf{r}^*$, transforming the non-linear projection into a tractable least-squares problem.  

The accuracy of this linearization hinges on the choice of $\mathbf{r}^*$, which must ensure that the first-order expansion closely aligns with the true non-linear behavior of the density function $\rho(\mathbf{k})$. However, selecting an optimal $\mathbf{r}^*$ is highly non-trivial, as it depends on the spatial structure of $\rho_0(\mathbf{k})$. Theoretically, this can be addressed by computing the cross-correlation between $\rho_0$ and $\rho(k;r)$, on a discretized grid to quantify their alignment.  While theoretically sound, this approach faces two practical limitations, 
\begin{enumerate}
    \item High computational cost, since correlations must be evaluated across all candidate $\mathbf{r}^*$ values,  
    \item Dependence on real-space discretization, which can obscure fine features in $\rho_0(\mathbf{k})$.  
\end{enumerate}

To overcome these challenges, we compute the quasi-projection through an iterative tangent-space approximation of the distance function around a reference point given by the current best estimate of the projection. This approach begins with an initial $\mathbf{r}^*$ chosen to minimize the positional component of the distance function. At each iteration, the density function is linearized about the current $\mathbf{r}^*$, and the resulting least-squares problem is solved to obtain a projection to the tangent space. The position and density variables are updated accordingly, and the procedure is repeated until convergence.

The initial choice of $\mathbf{r}^*$ is given by the weighted average of the position replicas:
\begin{align}\label{eq:average_position}
    \mathbf{r}^* &= P\big(\mathbf{r}^{(0)}\big) 
    = \arg\min_{\mathbf{r}} \sum_{d \in \mathcal{D}} \eta_d^2 \norm{\mathbf{r} - \mathbf{r}^{(0)}_d}^2, \\
    \mathbf{r}^* &= \frac{\sum\limits_{d \in \mathcal{D}} \eta_d^2 \mathbf{r}^{(0)}_d}{\sum\limits_{d \in \mathcal{D}} \eta_d^2}.
\end{align}

The corresponding squared-distance function in the tangent space approximation is
\begin{align}\label{eq:distance_tangent_space}
    \Delta_T^2\Big(\boldsymbol{\delta r}; \mathbf{r}^{(0)}, \rho^{(0)}(\mathbf{k})\Big) 
    &= \sum_{d \in \mathcal{D}} \eta_d^2 \norm{\mathbf{r}^* + \delta \mathbf{r} - \mathbf{r}^{(0)}_d}^2 \nonumber \ + \\
    &\quad g^2 \sum_{\mathbf{k} \in \mathbf{K}} \eta_{\mathbf{k}}^2 \norm{\rho(\mathbf{k}; 0) 
    \exp(i\mathbf{k} \cdot \mathbf{r}^*)\left(1 + i\mathbf{k} \cdot \boldsymbol{\delta r}\right) 
    - \rho^{(0)}(\mathbf{k})}^2.
\end{align}

Projection onto the tangent space for a reference point $\mathbf{r}^*$ is obtained by minimizing $\Delta_T^2$:  
\begin{align}\label{eq:r_T}
    \mathbf{r}_T = \mathbf{r}^* + \arg\min_{\boldsymbol{\delta r}} \Delta_T^2\Big(\boldsymbol{\delta r}; \mathbf{r}^{(0)}, \rho^{(0)}(\mathbf{k})\Big)
\end{align}
The distance-minimizing $\delta \mathbf{r}$ is determined by setting the gradient of $\Delta_T^2(\boldsymbol{\delta r}; \mathbf{r}^{(0)}, \rho^{(0)}(\mathbf{k}))$ to zero,
\begin{align}\label{eq:delta_r_tangent}
    \nabla_{\boldsymbol{\delta r}} \Delta_T^2 &= 0, \nonumber \\
    \implies \boldsymbol{\delta r} \left(\sum_{d \in \mathcal{D}} \eta_d^2\right) &+ g^2 \sum_{\mathbf{k} \in \mathbf{K}} \eta_{\mathbf{k}}^2 \Biggl( \rho^2(\mathbf{k}; 0) (\mathbf{k} \cdot \boldsymbol{\delta r}) \mathbf{k} \nonumber \\
    &- \text{Re}\left( i\mathbf{k} \, \rho(\mathbf{k}; 0) \exp(i\mathbf{k} \cdot \mathbf{r}^*) \left(\rho^{(0)}(\mathbf{k})\right)^\dagger \right) \Biggr) = 0.
\end{align}
This linearized system can be expressed compactly in matrix form,
\begin{align}\label{eq:matrix_equation_tangent}
    \mathbf{M} \boldsymbol{\delta r} = \mathbf{C},
\end{align}
with matrix $\mathbf{M}$ and vector $\mathbf{C}$ defined as
\begin{align}\label{eq:MC_tangent}
    \mathbf{M} &= \Big(\sum_{d \in \mathcal{D}} \eta_d^2 \Big) \mathbf{I} 
    + g^2 \sum_{\mathbf{k} \in \mathbf{K}} \eta_{\mathbf{k}}^2 \rho^2(\mathbf{k}; 0) \mathbf{k}\mathbf{k}^{\intercal}, \\
    \mathbf{C} &= g^2 \sum_{\mathbf{k} \in \mathbf{K}} \eta_{\mathbf{k}}^2 \;
    \text{Re}\Big( i\mathbf{k} \, \rho(\mathbf{k}; 0) \exp(i\mathbf{k} \cdot \mathbf{r}^*) \big(\rho^{(0)}(\mathbf{k})\big)^\dagger \Big),
\end{align}
where $\mathbf{I}$ is the identity matrix and $\mathbf{k}\mathbf{k}^T$ denotes the outer product of a scattering vector with itself, yielding a rank-1 positive semi-definite matrix.

The iterative tangent-space scheme is initialized with $P_0 = (\mathbf{r}^*, \boldsymbol{\rho}_0)$, followed by successive tangent-space projections:
\[
    P_1 = Q(P_0) = \left( \mathbf{r}^* + \boldsymbol{\delta r},\; 
    \left( \rho(\mathbf{k}; 0) \exp(i\mathbf{k} \cdot \mathbf{r}^*) 
    \left(1 + i\mathbf{k} \cdot \boldsymbol{\delta r} \right) \right)_{\mathbf{k} \in \mathbf{K}} \right).
\]
By repeated application, a sequence $P_n = Q^n(P_0)$ is generated, which converges rapidly to the concur constraint set. In practice, convergence is typically observed within 2--3 iterations, after which atomic position updates become negligible ($\|\boldsymbol{\delta r}\| < 10^{-6}$), demonstrating the efficient convergence by this method.  
\subsection{Effect of Space Group Symmetry}

Protein crystals often form in repeating lattice arrangements dictated by space group symmetry. For the crystal studied in this work, the space group is $\mathrm{P2_1 2_1 2_1}$, which contains four symmetry-equivalent positions per asymmetric unit. In this setting, if an atom occupies coordinates $(x, y, z)$, three additional symmetry-related copies are generated by the action of
\begin{enumerate}
    \item $\left(-x + \tfrac{1}{2},\; -y,\; z + \tfrac{1}{2}\right)$, 
    \item $\left(-x,\; y + \tfrac{1}{2},\; -z + \tfrac{1}{2}\right)$, 
    \item $\left(x + \tfrac{1}{2},\; -y + \tfrac{1}{2},\; -z\right)$.
\end{enumerate}
These operations correspond to $2_1$ screw axes along each of the three crystallographic directions ($a$, $b$, and $c$), where a $180^\circ$ rotation is combined with a half-unit translation along the axis. Such symmetry introduces additional constraints by enforcing consistency between the electron density map and the corresponding atomic positions across all equivalent sites.  

To incorporate these constraints, we define a symmetry-augmented distance function,  
\begin{equation}\label{eq:distance_mapping_constraint_full}
\Delta_{\text{U}}^2(\mathbf{r}; \mathbf{r}^{(0)}, \rho^{(0)}(\mathbf{k})) 
= \underbrace{\sum_{d \in \mathcal{D}} \eta_d^2 \norm{\mathbf{r} - \mathbf{r}^{(0)}_d}^2 \nonumber}_{\text{positional}} 
+ g^2 \underbrace{\sum_{\substack{\mathbf{k} \in \mathbf{K} \\ \nu \in \{0,1,2,3\}}} \eta_{\mathbf{k}}^2 \norm{\rho(\mathbf{k}; 0) \exp(i\mathbf{k} \cdot \mathbf{r}_\nu) - \rho_{\nu}^{(0)}(\mathbf{k})}^2}_{\text{symmetry-resolved density}}.
\end{equation}
where:
\begin{itemize}
    \item $\nu$ indexes the four asymmetric units in the primitive cell,
    \item $\mathbf{r}_\nu$ denotes symmetry-generated positions derived from $\mathbf{r}_0$,
    \item $\rho_{\nu} ^{(0)}(\mathbf{k})$ represents the current density iterates for each asymmetric unit.
\end{itemize}

This formulation ensures that both positional and density terms are evaluated over all symmetry-related sites, thereby enforcing global consistency across the unit cell.  

The complete unit cell distance function is tangentially approximated similarly to equation \eqref{eq:delta_r_tangent}, yielding a quasi-projection to the concur constraint.
\begin{align}\label{eq:delta_r_unit}
    \nabla_{\boldsymbol{\delta r}} (\Delta_U)_T^2 &= 0, \nonumber \\
    \implies \boldsymbol{\delta r} \left(\sum_{d \in \mathcal{D}} \eta_d^2\right) &+ g^2 \sum_{\substack{\mathbf{k} \in \mathbf{K} \\ \nu \in \{0,1,2,3\}}}\eta_{\mathbf{k}}^2 \Biggl( \rho^2(\mathbf{k}; 0) (\mathbf{k_\nu} \cdot \boldsymbol{\delta r}) \mathbf{k_\nu} \nonumber \\
    &- \text{Re}\left( i\mathbf{k_\nu} \, \rho(\mathbf{k}; 0) \exp(i\mathbf{k} \cdot \mathbf{r_\nu}^*) \left(\rho_{\nu}^{(0)}(\mathbf{k})\right)^\dagger \right) \Biggr) = 0.
\end{align}
where, $\mathbf{k_\nu} \equiv \nabla_{\mathbf{r}}(\mathbf{k}\cdot \mathbf{r_\nu})$.

The linearized system in \eqref{eq:delta_r_unit} can be written in matrix form as  
\[
\mathbf{M_U} \, \boldsymbol{\delta r} = \mathbf{C_U},
\]  

where the symmetry operations of the $\mathrm{P2_1 2_1 2_1}$ space group cause the off-diagonal terms in $\mathbf{M_U}$ to cancel among themselves, leaving $\mathbf{M_U}$ as a diagonal matrix.

The matrix $\mathbf{M_U}$ and vector $\mathbf{C_U}$ are given by
\begin{align}\label{eq:MC_unit}
    \mathbf{M_U} &= \left(\sum_{d \in \mathcal{D}} \eta_d^2 \right) \mathbf{I} 
    + 4 g^2 \sum_{\mathbf{k} \in \mathbf{K}} \eta_{\mathbf{k}}^2 \rho^2(\mathbf{k}; 0) \operatorname{diag}(\mathbf{k} \odot \mathbf{k}), \\
    \mathbf{C_U} &= g^2 \sum_{\substack{\mathbf{k} \in \mathbf{K} \\ \nu \in \{0,1,2,3\}}} \eta_{\mathbf{k}}^2 
    \text{Re}\left( i\mathbf{k_\nu} \, \rho(\mathbf{k}; 0) \exp(i\mathbf{k} \cdot \mathbf{r_\nu}^*) \left(\rho_{\nu}^{(0)}(\mathbf{k})\right)^\dagger \right).
\end{align}

In addition to the tangent-space scheme, alternative optimization strategies can be considered. One such approach is the use of quasi-Newton updates, such as the regularized Newton method, which incorporate curvature information to accelerate convergence. In practice, however, this strategy did not provide tangible improvements over the iterative tangent-space approximation. Both methods exhibited similar convergence behavior, but the quasi-Newton approach was consistently more computationally expensive due to the need to evaluate and regularize Hessian-like terms.  

This observation highlights a central advantage of the tangent-space formulation: It achieves efficient convergence with minimal computational cost. By avoiding the overhead of higher-order updates, the method remains well-suited for large-scale crystallographic refinement problems, where repeated projections must be performed rapidly across thousands of atoms that incorporate thousands of scattering vectors.

\begin{acknowledgements}
We thank Robert Thorne for bringing this problem to our attention and James Holton for insightful discussions and for making a detailed report of this challenge available on his website.
\end{acknowledgements}

\ConflictsOfInterest{The authors have no conflicts to disclose.}

\DataAvailability{All code and data used in this work are available under \texttt{MIT} license at
\url{https://github.com/avinash-mandaiya/refinement-release}.}

\bibliography{iucr} 

\end{document}